\documentclass{appolb}
\usepackage{epsfig}

\begin{document}
\title{Coincidence of Universe age in $\Lambda$CDM and Milne cosmologies
}
\author{{Marek Kutschera$^{a,b}$\thanks{e-mail: Marek.Kutschera@ifj.edu.pl}, Micha\l { }Dyrda$^a$\thanks{e-mail: dyrda@th.if.uj.edu.pl}}
\address{~$^a$ The M. Smoluchowski Institute of Physics, Jagellonian University, Reymonta 4, 30-059 Krak\'ow, Poland}
\address{~$^b$ The H. Niewodnicza\'nski Institute of Nuclear Physics, Radzikowskiego 142, 31-342 Krak\'ow, Poland}
}

\maketitle
\begin{abstract}
The age of the Universe in the $\Lambda$CDM cosmology with $\Omega_{\textrm{matter}}=0.26$ and $\Omega_{\Lambda}=0.74$ is the same as in the Milne cosmology which correspods to an almost empty universe. In both cases it is a reciprocal Hubble constant, $1/H_0$, that for now preferred value $H_0=71\, \textrm{km/s/Mpc}$ is $13.7$ billion years. The most curious coincidence is that at the present time, in the $\Lambda$CDM model the decelerated expansion is exactly compensated by the accelerated expansion, as if the Universe coast for $13.7$ billion years.
\end{abstract}
\PACS{95.36.+x, 98.80.-k}

\bigskip

The current concordance cosmological model is a flat $\Lambda$CDM model with the matter fraction $\Omega_{\textrm{matter}}=0.26$ and the vacuum energy contribution $\Omega_{\Lambda}=0.74$. The baryon matter contribution, of about 4 percent, is included here in $\Omega_{\textrm{matter}}$. 

The age of the universe, $t_0$, is calculated for any cosmological model with the formula \cite{peebles}
\begin{equation}
H_0t_0 = \int_0^\infty \frac{dz}{(1+z)E(z)},
\label{eq1}
\end{equation}
where the function $E(z)$ reads:
\begin{equation}
E(z)=\left[\Omega_{\textrm{matter}}(1+z)^3+\Omega_R(1+z)^2+\Omega_{\Lambda} \right]^{1/2},
\end{equation}
and the mass parameters sum up to unity: $\Omega_{\textrm{matter}}+\Omega_R+\Omega_{\Lambda}=1$. Here $\Omega_R$ is the curvature contribution, which vanishes for flat cosmological models.

Consider first the Milne cosmology \cite{milne}. As matter content is negligible, $\Omega_{\textrm{matter}}\approx 0$, and there is no cosmological constant, $\Omega_{\Lambda}=0$, hence $\Omega_R=1$. The integral in eq.(\ref{eq1}) is equal to 1, and $H_0t_0=1$.

For a flat cosmology, $\Omega_R=0$, the age of the universe depends only on matter density parameter, $\Omega_{\textrm{matter}}$, as the vacuum energy is $\Omega_{\Lambda}=1-\Omega_{\textrm{matter}}$. Let us solve the equation 
\begin{equation}
\int_0^\infty \frac{dz}{(1+z)E(z)}=1
\end{equation}
for $\Omega_{\textrm{matter}}$.
One easily finds the solution (with required accuracy of two digits), $\Omega_{\textrm{matter}}=0.26$, and $\Omega_{\Lambda}=0.74$. These are precisely the values of mass parameters describing content of the Universe in recently released  three year results of WMAP collaboration \cite{spergel}.
The age of the Universe in the concordance model is thus the same as in the Milne cosmology, quite a striking coincidence.

 \begin{figure}
 \begin{center}
 \includegraphics[width=85mm]{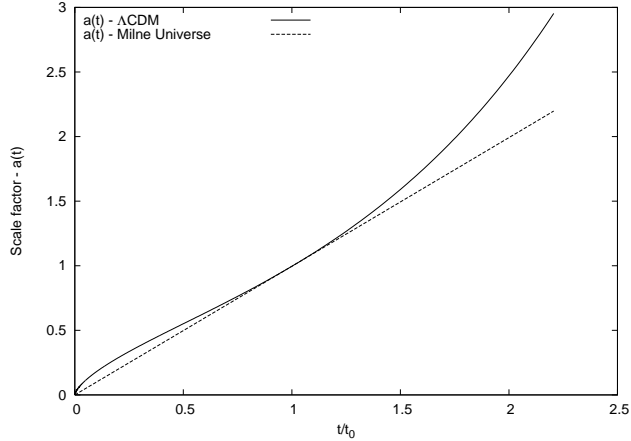}
 \caption{\label{fig1} Comparison of $\Lambda$CDM concordance cosmology with Milne cosmology. For $t/t_0$=1 and $a(t_0)=1$, where $t_0$ is the age of the Universe, the scale factor for Milne model is tangent to that of $\Lambda$CDM cosmological model.}
 \end{center}
 \end{figure}
 
In Fig.1 we show the scale factor of the Universe as a function of cosmic time for both models. As one can see, the two curves are tangent to one another only at a single point, namely at the present moment in the whole history of the Universe. Physically, this means that just now the influence of the accelerated expansion of the Universe exactly compensates the opposite influence of the initial, decelerated expansion, on any length scale.

This may be just an innocent coincidence, however, unexpected coincidences in cosmology require scrutiny. The one discussed here, that gravity and antigravity contributions cancel each other, could perhaps indicate that the gravity in actual universe is not of infinite range.

\bigskip

This research is partially supported by the Polish Ministry for Science and Informatization, grant no 1P03D00528.

\end{document}